\documentclass[conference, 10pt, final]{IEEEtran}

\usepackage{enumitem}
\usepackage[nolist]{acronym} 
\usepackage{tikz}
\usepackage{pgfplots}
\usepgfplotslibrary{groupplots}
\usepackage{graphicx}
\usepackage{float}
\pgfplotsset{compat=newest}
\usepackage{units}
\usepackage{amsmath, amsbsy, amssymb}
\usepackage{tikzscale}
\usetikzlibrary{arrows}
\usepackage{color}
\usepackage{epigraph}
\usepackage{circuitikz}
\usepackage{multirow}
\usepackage{booktabs}
\usepackage[plainruled]{algorithm2e}
\usepackage[group-separator={,}]{siunitx}
\definecolor{darkgreen}{rgb}{0.125,0.5,0.169}
\usetikzlibrary{shapes,arrows}
\usetikzlibrary{positioning}
\usetikzlibrary{patterns}
\usetikzlibrary{decorations.pathreplacing}
\usepackage{marvosym}
\usepackage{bbm}
\usepackage{mathtools}
\usepackage{xfrac}

\tikzset{>=latex}

\renewcommand{\vec}[1]{\mathbf{#1}}

\newcommand{\bv}{\vec{b}}

\newcommand{\lv}{\vec{l}}

\newcommand{\nv}{\vec{n}}

\newcommand{\xv}{\vec{x}}
\newcommand{\yv}{\vec{y}}

\newcommand{\Hm}{\vec{H}}

\newcommand{\CC}{\mathbb{C}}

\newcommand{\LB}{\left(}
\newcommand{\RB}{\right)}

\renewcommand{\ln}[1]{\mathop{\mathrm{ln}}\LB #1\RB}

\begin{acronym}
 \acro{CSI}{channel state information}
 \acro{UE}{user equipment}
 \acro{UL}{uplink}
 \acro{BS}{basestation}
 \acro{TDD}{time division duplex}
 \acro{FDD}{frequency division duplex}
 \acro{ECC}{error-correcting code}
 \acro{MLD}{maximum likelihood decoding}
 \acro{HDD}{hard decision decoding}
 \acro{IF}{intermediate frequency}
 \acro{RF}{radio frequency}
 \acro{SDD}{soft decision decoding}
 \acro{NND}{neural network decoding}
 \acro{CNN}{convolutional neural network}
 \acro{ML}{maximum likelihood}
 \acro{GPU}{graphical processing unit}
 \acro{BP}{belief propagation}
 \acro{LTE}{Long Term Evolution}
 \acro{BER}{bit error rate}
 \acro{SNR}{signal-to-noise-ratio}
 \acro{ReLU}{rectified linear unit}
 \acro{BPSK}{binary phase shift keying}
 \acro{QPSK}{quadrature phase shift keying}
 \acro{AWGN}{additive white Gaussian noise}
 \acro{MSE}{mean squared error}
 \acro{LLR}{log-likelihood ratio}
 \acro{MAP}{maximum a posteriori}
 \acro{NVE}{normalized validation error}
 \acro{BCE}{binary cross-entropy}
 \acro{CE}{cross-entropy}
 \acro{BLER}{block error rate}
 \acro{SQR}{signal-to-quantisation-noise-ratio}
 \acro{MIMO}{multiple-input multiple-output}
 \acro{OFDM}{orthogonal frequency division multiplex}
 \acro{RF}{radio frequency}
 \acro{LOS}{line of sight}
 \acro{NLoS}{non-line of sight}
 \acro{NMSE}{normalized mean squared error}
 \acro{CFO}{carrier frequency offset}
 \acro{SFO}{sampling frequency offset}
 \acro{IPS}{indoor positioning system}
 \acro{TRIPS}{time-reversal IPS}
 \acro{RSSI}{received signal strength indicator}
 \acro{MIMO}{multiple-input multiple-output}
 \acro{ENoB}{effective number of bits}
 \acro{AGC}{automatic gain control}
 \acro{ADC}{analog to digital converter}
 \acro{ADCs}{analog to digital converters}
 \acro{FB}{front bandpass}
 \acro{FPGA}{field programmable gate array}
 \acro{JSDM}{Joint Spatial Division and Multiplexing}
 \acro{NN}{neural network}
 \acro{IF}{intermediate frequency}
 \acro{LoS}{line-of-sight}
 \acro{NLoS}{non-line-of-sight}
 \acro{DSP}{digital signal processing}
 \acro{AFE}{analog front end}
 \acro{SQNR}{signal-to-quantisation-noise-ratio}
 \acro{SINR}{signal-to-interference-noise-ratio}
 \acro{ENoB}{effective number of bits}
 \acro{PCB}{printed circuit board}
 \acro{EVM}{error vector mangnitude}
 \acro{CDF}{cumulative distribution function}
 \acro{MRC}{maximum ratio combining}
 \acro{MRP}{maximum ratio precoding}
 \acro{MRT}{maximum ratio transmission}
 \acro{DeepL}{deep-learning}
 \acro{DL}{deep learning}
 \acro{SISO}{single-input single-output}
 \acro{SGD}{stochastic gradient descent}
 \acro{CP}{cyclic prefix}
 \acro{MISO}{Multiple Input Single Output}
 \acro{LMMSE}{linear minimum mean square error}
 \acro{ZF}{zero forcing}
 \acro{USRP}{universal software radio peripheral}
 \acro{RNN}{recurrent neural network}
 \acro{GRU}{gated recurrent unit}
 \acro{LSTM}{long short-term memory}
 \acro{NTM}{neural turing machine}
 \acro{DNC}{differentiable neural computer}
 \acro{TCN}{temporal convolutional network}
 \acro{FCL}{fully connected layer}
 \acro{MANN}{memory augmented neural network}
 \acro{RNN}{recurrent neural network}
 \acro{DNN}{dense neural network}
 \acro{FIR}{finite impulse response}
 \acro{BPTT}{back-propagation through time}
 \acro{GAN}{generative adversarial network}
 \acro{ELU}{exponential linear unit}
 \acro{tanh}{hyperbolic tangent}
 \acro{BICM}{bit-interleaved coded modulation}
 \acro{OTA}{over-the-air}
 \acro{IM}{intensity modulation}
 \acro{DD}{direct detection}
 \acro{RL}{reinforcement learning}
 \acro{SDR}{software-defined radio}
 \acro{WGAN}{Wasserstein generative adversarial network}
 \acro{BMD}{bit-metric decoding}
 \acro{BMI}{bit-wise mutual information}
 \acro{LDPC}{low-density parity-check}
 \acro{IDD}{iterative demapping and decoding}
 \acro{JSD}{Jensen-Shannon divergence}
 \acro{MMSE}{minimum mean square error}
 \acro{FFT}{fast Fourier transform}
 \acro{IFFT}{inverse fast Fourier transform}
 \acro{QAM}{quadrature amplitude modulation}
 \acro{EMD}{earth mover's distance}
 \acro{TDL}{tapped delay line}
 \acro{KL}{Kullback-Leibler}
 \acro{PRACH}{physical random access channel}
 \acro{URLLC}{ultra-reliable low-latency communication}
 \acro{ANOMA}{asynchronous non-orthogonal multiple access}
 \acro{FEC}{forward error correction}
 \acro{PAPR}{peak-to-average power ratio}
 \acro{APP}{a posteriori probability}
 \acro{COTS}{commercial off-the-shelf}
 \acro{PLL}{phase locked loop}
 \acro{STO}{sampling time offset}
 \acro{SFO}{sampling frequency offset}
 \acro{CFO}{carrier frequency offset}
 \acro{CPO}{carrier phase offset}
 \acro{CSI}{channel state information}
 \acro{GNSS}{global navigation satellite system}
 \acro{ELAA}{extremely large aperture array}
 \acro{UE}{user equipment}
 \acro{DICHASUS}{\underline{Di}stributed \underline{Cha}nnel \underline{S}ounder by \underline{U}niversity of \underline{S}tuttgart}
 \acro{JCAS}{Joint Communication and Sensing}
\end{acronym}

\definecolor{mittelblau}{RGB}{0, 126, 198}
\definecolor{violettblau}{cmyk}{0.9, 0.6, 0, 0}
\definecolor{rot}{RGB}{238, 28 35}
\definecolor{apfelgruen}{RGB}{140, 198, 62}
\definecolor{gelb}{RGB}{1, 221, 0}
\definecolor{orange}{RGB}{244, 111, 33}
\definecolor{pink}{RGB}{237, 0, 140}
\definecolor{lila}{RGB}{128, 10, 145}
\definecolor{hellgrau}{RGB}{224, 224, 224}
\definecolor{mittelgrau}{RGB}{128, 128, 128}
\definecolor{dunkelgrau}{RGB}{80,80,80}
\definecolor{anthrazit}{RGB}{19, 31, 31}

\pgfplotscreateplotcyclelist{corporate colours markers}{%
rot, every mark/.append style={fill=.!80!rot},mark=*\\%
mittelblau, every mark/.append style={fill=.!80!mittelblau},mark=square*\\%
apfelgruen, every mark/.append style={fill=.!80!apfelgruen},mark=triangle*\\%
orange, mark=star\\%
pink, every mark/.append style={fill=.!80!pink},mark=diamond*\\%
violettblau, every mark/.append style={fill=.!80!violettblau},mark=otimes*\\%
lila, mark=|\\%
gelb, every mark/.append style={fill=.!80!gelb},mark=pentagon*\\%
hellgrau, mark=text,text mark=p\\%
anthrazit, mark=text,text mark=a\\%
}

\pgfplotscreateplotcyclelist{corporate colours markers double}{%
rot, every mark/.append style={fill=.!80!rot},mark=*\\%
rot, dashed, every mark/.append style={fill=.!80!rot,solid},mark=*\\%
mittelblau, every mark/.append style={fill=.!80!mittelblau},mark=square*\\%
mittelblau, dashed, every mark/.append style={fill=.!80!mittelblau,solid},mark=square*\\%
apfelgruen, every mark/.append style={fill=.!80!apfelgruen},mark=triangle*\\%
apfelgruen, dashed, every mark/.append style={fill=.!80!apfelgruen,solid},mark=triangle*\\%
orange, mark=star\\%
orange, dashed, mark=star\\%
pink, every mark/.append style={fill=.!80!pink},mark=diamond*\\%
pink, dashed, every mark/.append style={fill=.!80!pink,solid},mark=diamond*\\%
violettblau, every mark/.append style={fill=.!80!violettblau},mark=otimes*\\%
violettblau, dashed, every mark/.append style={fill=.!80!violettblau,solid},mark=otimes*\\%
lila, mark=|\\%
lila, dashed, mark=|\\%
gelb, every mark/.append style={fill=.!80!gelb},mark=pentagon*\\%
gelb, dashed, every mark/.append style={fill=.!80!gelb,solid},mark=pentagon*\\%
}

\definecolor{bgorange}{HTML}{fcc0a7}
\definecolor{bggreen}{HTML}{ccebb9}

\IEEEoverridecommandlockouts

\begin{document}

\title{A Distributed Massive MIMO Channel Sounder for ``Big CSI Data''-driven Machine Learning} 

\author{\IEEEauthorblockN{Florian Euchner, Marc Gauger, Sebastian D\"orner, Stephan ten Brink \\}

\IEEEauthorblockA{
Institute of Telecommunications, Pfaffenwaldring 47, University of  Stuttgart, 70569 Stuttgart, Germany \\ \{euchner,gauger,doerner,tenbrink\}@inue.uni-stuttgart.de
}

}

\maketitle

\begin{abstract}
A distributed massive \ac{MIMO} channel sounder for acquiring large \ac{CSI} datasets, dubbed \ac{DICHASUS}, is presented.
The measured data has potential applications in the study of various machine learning algorithms for user localization, \ac{JCAS}, channel charting, enabling massive MIMO in \ac{FDD} operation,  and many others.
The proposed channel sounder architecture is distinct from similar previous designs in that each individual single-antenna receiver is completely autonomous, enabling arbitrary, spatially distributed antenna deployments, and offering virtually unlimited scalability in the number of antennas.
Optionally, extracted channel coefficient vectors can be tagged with ground truth position data, obtained either through a \ac{GNSS} receiver (for outdoor operation) or through various indoor positioning techniques.
\end{abstract}

\acresetall

\section{Introduction}
To study massive \ac{MIMO} channels, accurate wireless channel measurements from a multitude of receive antennas are required.
Such channel measurements, referred to as \ac{CSI}, are acquired through a process known as channel sounding.
Ever since the concept of massive \ac{MIMO} was first introduced, the scientific community has placed significant effort into obtaining reliable and large \ac{CSI} datasets.
Early massive \ac{MIMO} channel sounding experiments have been carried out since 2012 \cite{hekaton} \cite{argos} \cite{firstmami}, followed by many more advanced channel sounders developed by different research groups in the subsequent years \cite{bristol} \cite{lund} \cite{euroecom}.
Nowadays, real-time massive \ac{MIMO} testbeds have become commercially available \cite{nutaq} \cite{ni} \cite{skylarkwireless}, albeit at a rather expensive price point.
More recently, an open source initiative for massive \ac{MIMO} development, with a channel sounder based on hardware components from Skylark Wireless \cite{skylarkwireless}, was launched by a research group at Rice University \cite{renew}.
All of these channel sounders, including a previous channel sounder developed at our institute \cite{arnold2019novel}, are primarily targeted at centralized massive \ac{MIMO} deployments where all receivers are concentrated in a single antenna array.

While the idea of cell-less wireless networks is not new, in combination with massive \ac{MIMO}, the concept has seen a surge in research interest over the last years.
Research results, though only of theoretical nature, suggest that cell-less massive MIMO, also known as distributed massive \ac{MIMO}, has advantages over network architectures based on small cells \cite{ngo2017cell}.
One objective of our channel sounder, \ac{DICHASUS}, is to facilitate research into various distributed antenna deployment schemes, such as \acp{ELAA}, where antennas are distributed along building facades or embedded into building structures as holographic antennas, as well as truly distributed antenna deployments across multiple buildings.

Therefore, \ac{DICHASUS} was designed with a decentralized system architecture in mind:
Rather than collecting all measurements at a central node, each receiver acquires and stores the samples from each antenna autonomously while staying tightly synchronized to a reference signal transmitter.
This over-the-air synchronization method not only provides the flexibility to build distributed setups, but also enables straightforward scalability in the number of antennas and reduces overall system set-up complexity, which allows for rapid deployments of the channel sounder.
A breakdown of the system architecture and a simple verification experiment are presented in Sections \ref{sec:overview} through \ref{sec:verification}.

\begin{figure}
    \includegraphics[width=\columnwidth]{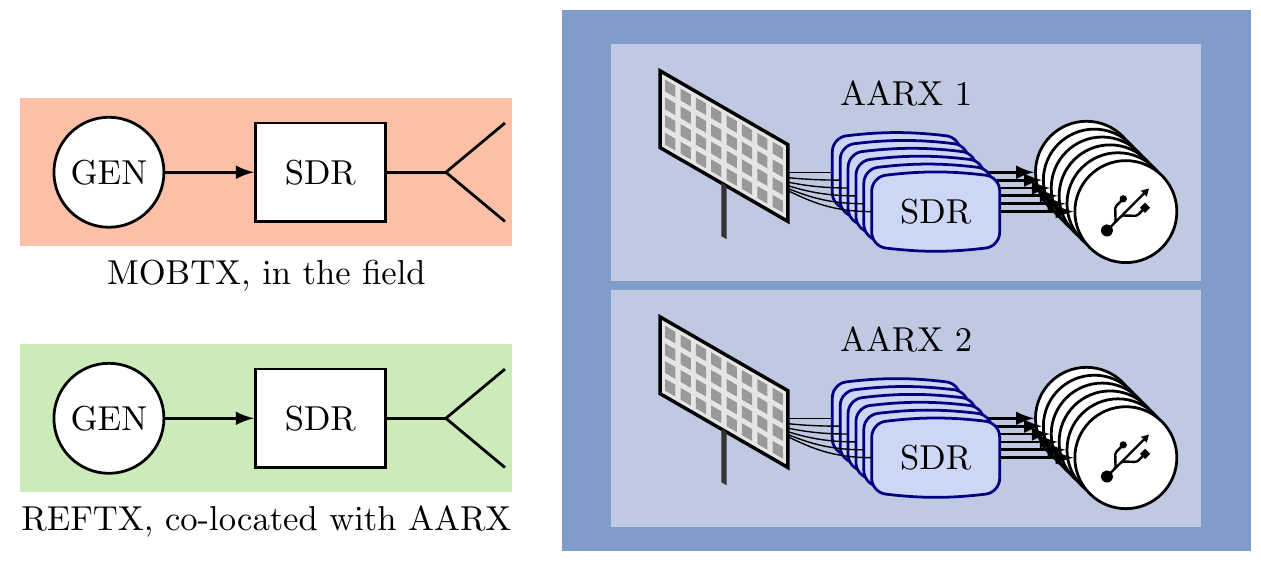}
    \caption{Massive \ac{MIMO} system set-up with single-antenna, mobile transmitter (MOBTX), reference transmitter (REFTX) and one or multiple individual antennas or antenna arrays with receivers (AARX) and USB flash storage.}
    \label{fig:system}
    \vspace{-0.5cm}
\end{figure}

Captured \ac{DICHASUS} datasets, one of which is described in Section \ref{sec:dataset}, are made freely available to the scientific community\footnote{Datasets, tutorials and sample code can be downloaded from \mbox{\texttt{https://dichasus.inue.uni-stuttgart.de}}.}.
In addition to the acquired \ac{CSI}, we provide accurate labels such as position tags.
In Section \ref{sec:nn}, we demonstrate one possible application of the captured \ac{CSI} data, namely \ac{NN}-based localization.

\section{Overall System Set-up}
\label{sec:overview}

\ac{DICHASUS} covers the frequency range from 70 MHz to 6 GHz and every receiver has an \ac{ADC} bit depth of $12\,\mathrm{bit}$.
As depicted in Fig.~\ref{fig:system}, it consists of 
\begin{enumerate}[label=\alph*)]
    \item one or several groups of antenna arrays (AARX) as receivers, mimicking distributed base stations, described in detail in Section \ref{sec:aarx},
    \item a reference transmitter (REFTX) for frequency, phase and timing calibration, at a location that ensures a low-attenuation propagation path to all AARX antennas, explained in Section \ref{sec:reftx} and
    \item a mobile transmitter (MOBTX) with one transmit antenna in the field, which resembles a \ac{UE}, described in Section \ref{sec:mobtx}.
\end{enumerate}
The MOBTX continuously transmits \ac{CP}-\ac{OFDM}-modulated pilot symbols with a configurable number of subcarriers $N_\mathrm{sub, mob}$ and a maximal bandwidth of 50 MHz, enabling the extraction of \ac{CSI} per subcarrier at the receiver antenna array.
Also, the transmitter can embed its position, e.g., \ac{GNSS} coordinates, into the transmit signal, making it possible to obtain \ac{CSI} data together with position tags, which can later be used for supervised machine learning.
The REFTX is placed in such a way that there is a good propagation path to each AARX receive antenna, e.g., a direct line-of-sight path.
It also transmits \ac{CP}-\ac{OFDM}-based pilot signals with $N_\mathrm{sub, ref} = 128$ subcarriers at a bandwidth of $2\,\mathrm{MHz}$, but in a different, non-overlapping frequency range, that is, MOBTX and REFTX are frequency-multiplexed.
At all times, the AARX can receive both the signal from the MOBTX (through the channel of interest) as well as the signal from the REFTX.
The individual single-antenna AARX receivers operate in an autonomous fashion, i.e., without a central clock, but with clock signals that are purely derived from each receiver's built-in crystal oscillator.
Therefore, they rely on regular over-the-air frequency, phase and timing calibration using the REFTX's transmit signal, called REFTX broadcast.
While the REFTX frequency may be inaccurate in absolute terms due to its own oscillator drift, by synchronizing every receiver to the same frequency reference, the relative offsets between receivers disappear.

\section{Massive MIMO Receiver (AARX)}
\label{sec:aarx}
\begin{figure}
    \centering
    \includegraphics[width=0.8\columnwidth]{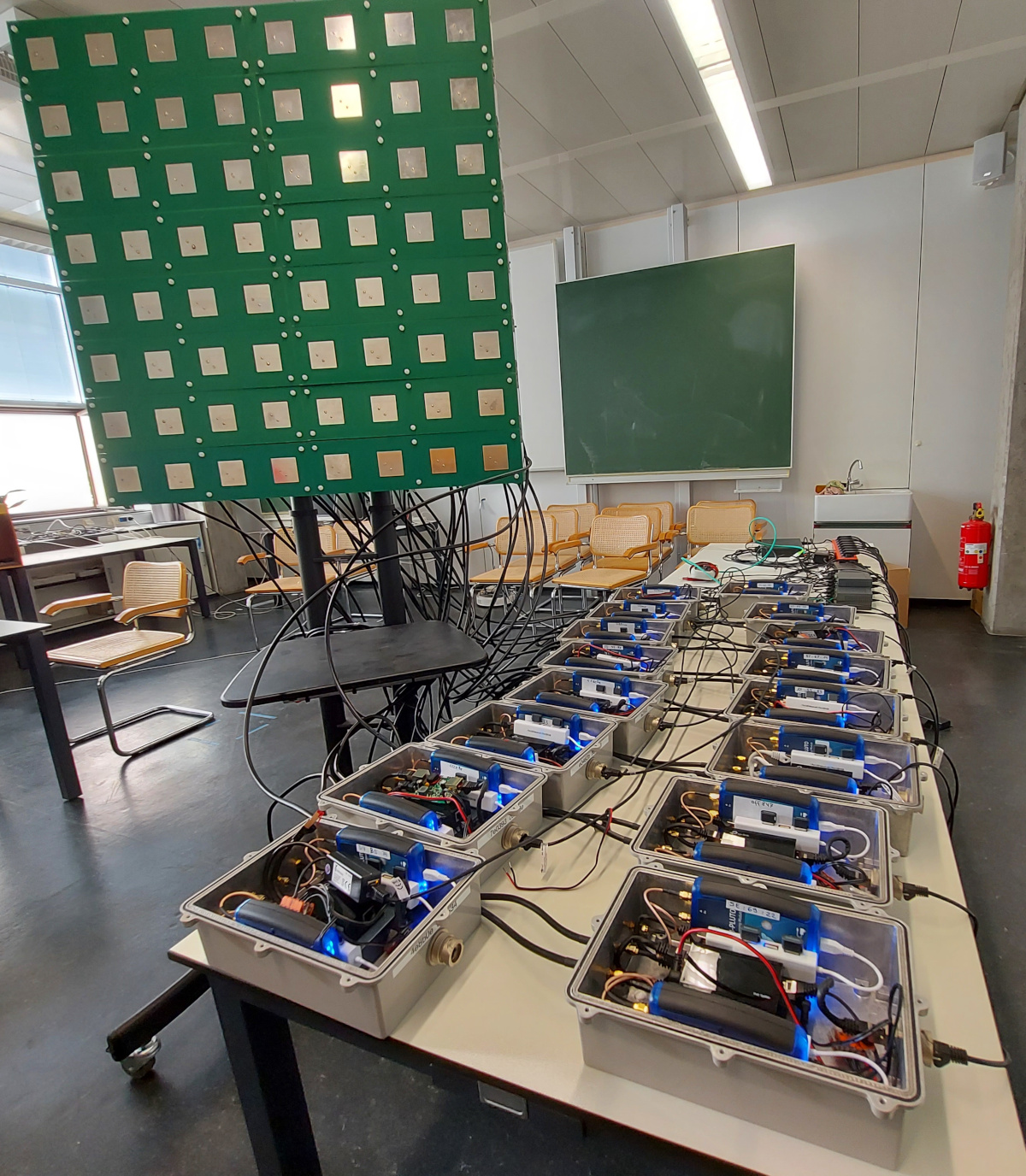}
    \caption{Photo of 64-antenna AARX array (only 32 antennas of which are connected) on the left and autonomous AARX receivers on the right.}
    \label{fig:mimo-rx}
    \vspace{-0.5cm}
\end{figure}

The AARX, shown in Fig. \ref{fig:mimo-rx}, is entirely built from comparably inexpensive \ac{COTS} hardware components.
Central to the AARX are Analog Devices' ADALM-PLUTO \acp{SDR} \cite{adalmpluto} that capture raw I/Q samples in a configurable frequency band received by the antennas.
Each ADALM-PLUTO contains a Cortex-A9 microprocessor running a Linux-based firmware, on top of which a custom application is responsible for managing the sample capture.
This way, all real-time signal processing functions can be performed by the receiver itself.
Captured I/Q data is first transferred to the device's RAM, where $200\,\mathrm{MB}$ per device are allocated for sample buffering, and later stored onto one or multiple USB storage devices attached to each ADALM-PLUTO.
While each ADALM-PLUTO is capable of capturing samples with a bandwidth of over $50\,\mathrm{MHz}$ to RAM, tests with typical USB storage devices show that a continuous writing speed of up to $15\,\sfrac{\mathrm{MB}}{\mathrm s}$ is realistic.
Real-time 12-bit I/Q sample capture is hence possible for capture rates of up to approximately $5\,\mathrm{MHz}$.
By, instead, only capturing short bursts, i.e., operating with a certain duty cycle, higher bandwidths are achievable.
For example, even for a capture bandwidth of $50\,\mathrm{MHz}$, the RAM buffer can hold more than one second of continuous I/Q samples.


All receivers are managed by a central controller, connected through Ethernet.
The central controller does not perform any signal processing, but provides a graphical user interface through which all parameters of the data capture can be configured.
Once the process of automatic calibration and capturing has been started, however, individual receivers can operate autonomously thanks to the reference transmitter (REFTX) concept, making this architecture easily scalable in the number of deployed AARX elements.
In fact, since receivers store data in a decentralized fashion on their own USB flash storage devices, there is no need for high-speed networking for collecting data at a central node.
During the channel sounding operation, the user interface on the central controller can be used for monitoring system parameters such as each antenna's \ac{SNR}, buffer state or storage write speed.

Within one measurement campaign, all recordings are performed with fixed gain configurations at all transmitters and receivers so that channel attenuations are comparable between measurements at different points in time and space, i.e., \ac{AGC} is disabled.

\section{REFTX and Channel Model}
\label{sec:reftx}
\begin{figure}
    \centering
    \begin{tikzpicture}
    \draw [-latex] (-0.3, 0) -- (8, 0) node[pos = 0.9, below = 0.15cm, anchor = north, xshift = 0.2cm] {time [ms]};
    
    \foreach [evaluate = \i as \ms using int(\i * 100)] \i in {0, 1, ..., 6} {
        \node [anchor = north, below = 0.15cm] at (\i, 0) {\ms};
        \draw (\i, -0.08) -- (\i, 0.08);
    }
    
    \draw [thick, fill = gray!30!white] (0, 0) rectangle (0.5, 1) node [midway, rotate = 90] {sync};
    
    \draw [thick, fill = gray!30!white] (6, 0) rectangle (6.5, 1) node [midway, rotate = 90] {sync};
    
    \draw [thick] (1, 0) rectangle (2, 1) node [midway, rotate = 90, align = center] {I/Q\\capt.};
    \draw [thick] (3, 0) rectangle (4, 1) node [midway, rotate = 90, align = center] {I/Q\\capt.};
    \draw [thick] (5, 0) rectangle (6, 1) node [midway, rotate = 90, align = center] {I/Q\\capt.};

    \node at (-0.35, 0.5) {$\cdots$};
    \node at (6.85, 0.5) {$\cdots$};
\end{tikzpicture}
    \caption{Typical frequency / timing synchronization and I/Q sample capture burst schedule}
    \label{fig:schedule}
    \vspace{-0.5cm}
\end{figure}
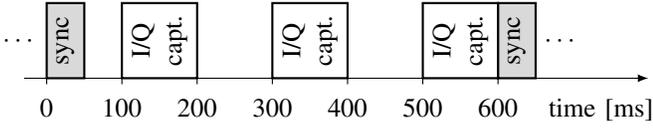

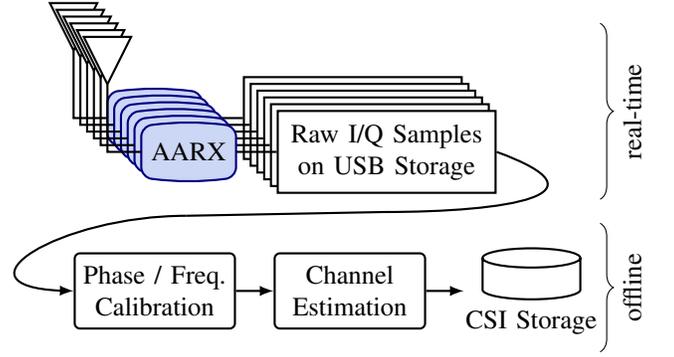
\begin{figure}
\newcommand{\plutosdr}[4] {
	\begin{scope}[shift = {#3}, scale = #2 / 11.8, rotate = #4]
	    \coordinate (#1-center) at (0, 0);
		\draw [thick, fill = blue!80!green!20!white, draw = blue!50!black, rounded corners] (-5.9, -3.2) to [bend right = 7] node[midway, inner sep = 0pt] (#1-south) {} (5.9, -3.2) to (5.9, 3.2) to [bend right = 7] node[midway, inner sep = 0pt] (#1-north) {} (-5.9, 3.2) -- cycle;
		\coordinate (#1-east) at (5.9, 0);
		\coordinate (#1-west) at (-5.9, 0);
	\end{scope}
}


\begin{circuitikz}
    \begin{scope}[shift = {(0, 0)}, scale = 0.9]
        \foreach \i [evaluate=\i as \x using \i * 0.1, evaluate=\i as \y using -\i * 0.1] in {0,...,5}{
            \plutosdr{pluto\i}{1.4}{(\x, \y))}{0};
            \draw  [thick] (pluto\i-east) -- +(0.6, 0) node [fill = white, anchor = west, draw, inner sep = 5pt, align = center] (usb\i) {Raw I/Q Samples\\on USB Storage};
            \draw [thick] (pluto\i-west) -- +(-0.5, 0) -- +(-0.5, 1) coordinate (antennabase\i);
            \draw [thick, fill = white] (antennabase\i) -- ($(antennabase\i) + (0.35, 0.7)$) -- ($(antennabase\i) + (-0.35, 0.7)$) -- cycle;
        }
        \node at (pluto5-center) {AARX};
        
        \draw [decorate,decoration={brace, amplitude = 5pt, mirror, raise = 2pt}, yshift=0pt]
(6.5, -1.2) -- (6.5, 1.4) node [black, midway, xshift=0.5cm, rotate = 90] {real-time};
    \end{scope}
    
    \tikzset{ppstep/.style={draw, thick, rounded corners = 2pt, minimum width = 2cm, minimum height = 1cm, align = center}}

    \begin{scope}[shift = {(0, -2.3)}, scale = 0.9]
        \node (sync) [ppstep] at (0, 0) {Phase / Freq.\\Calibration};
        \node (estim) [ppstep, right = 0.5cm of sync] {Channel\\ Estimation};
        
        \tikzset{database/.style={cylinder, inner sep = 4pt, aspect=0.7, minimum width = 1.0cm, draw, rotate=90, path picture={}}}
        
        \node (storage) [database, right = 1.4cm of estim, scale=1.3, thick, xshift = -0.1cm] {};
        \node [anchor = north] at (storage.west) {CSI Storage};
        
        \draw [decorate,decoration={brace, amplitude = 5pt, mirror, raise = 2pt}, yshift=0pt]
(6.5, -0.9) -- (6.5, 1.0) node [black, midway, xshift=0.5cm, rotate = 90] {offline};
    \end{scope}
    
    \coordinate (conne) at ($(usb5.east) + (-1.5, -0.8)$);
    \coordinate (connw) at ($(sync.west) + (1.5, 1)$);
    
    \draw  [thick] (usb5.east) edge[in = 0, out = -20, looseness = 2.5] (conne);
    \draw [thick] (connw) edge[-latex, in = 180, out = 180, looseness = 2.5] (sync.west);
    \draw [thick] (conne) -- (connw);
    \draw [thick, -latex] (sync) -- (estim);
    \draw [thick, -latex] (estim.east) -- +(0.5, 0);
\end{circuitikz}
    \caption{Massive \ac{MIMO} receiver architecture: Distributed sampling and raw data storage, centralized synchronization and analysis.}
    \label{fig:rx}
    \vspace{-0.5cm}
\end{figure}

With \ac{DICHASUS}, multiple receivers can be deployed in a spatially distributed manner as long as reception of the REFTX broadcast signal is ensured.
The REFTX signal is used for synchronization both in \emph{real time} and later in an \emph{off-line} postprocessing step, compare Fig. \ref{fig:rx}.
In real time, the REFTX broadcast is used for two aspects of synchronization:
\begin{itemize}
    \item Timing synchronization: The receivers typically do not record continuously, but in bursts, following a predefined schedule such as the one shown in Fig. \ref{fig:schedule}. All receivers align this schedule to the received REFTX broadcast in time, with an accuracy that corresponds to the duration of one I/Q sample.
    \item Frequency synchronization: Each receiver's sampling and local oscillator clock are derived from an internal crystal oscillator through a \ac{PLL}. By adjusting the \ac{PLL}'s fractional-$n$ divider, local oscillator and sampling clock frequencies are synchronized with the REFTX down to approximately 0.025ppm.
\end{itemize}

Compared to off-line frequency correction, real time \ac{CFO} and \ac{SFO} correction has the advantage that \ac{SFO}, the correction of which in digital processing is computationally expensive, becomes negligible over the short capture burst duration.
The subsequent off-line postprocessing step improves synchronization by performing even more timing offset corrections, namely:
\begin{itemize}
    \item \ac{CFO} correction down to significantly less than 0.025ppm
    \item Phase correction so that the local oscillators of all receivers after correction can be considered to be phase-synchronous, except for the phase shift of the mostly flat channel between REFTX and receiver.
    \item Sub-sample \ac{STO} correction so that the sampling instances can be considered to be identical across all receivers, except for the delay of the channel between REFTX and receiver.
\end{itemize}
This off-line synchronization is only possible due to the fact that the AARX receivers capture and store raw I/Q samples for the complete capture bandwidth, i.e., containing both REFTX and MOBTX signals.

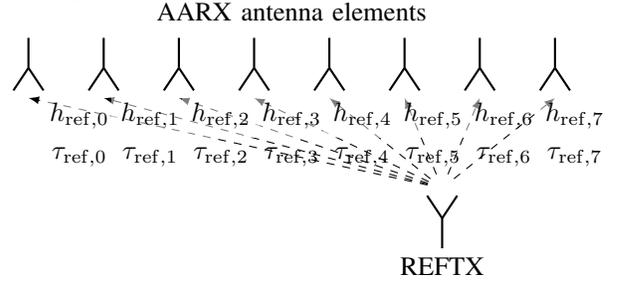
\begin{figure}
    \centering
    \begin{tikzpicture}
    \begin{scope}[shift = {(5.5, -2)}]
        \draw [thick] (0, 0) -- (0, -0.4);
        \draw [thick] (0, 0) -- (-0.2, 0.3);
        \draw [thick] (0, 0) -- ( 0.2, 0.3);
        \coordinate (reftx) at (0, 0.4) {};
        \node [anchor = north] at (0, -0.4) {REFTX};
    \end{scope}

    \foreach [evaluate = \i as \x using int(\i * 1), evaluate = \i as \xs using int(\i * 4 - 12)] \i in {0, 1, ..., 7} {
        \draw [thick] (\x, 0) -- (\x, 0.4);
        \draw [thick] (\x, 0) -- +(-0.2, -0.3);
        \draw [thick] (\x, 0) -- +( 0.2, -0.3);
        
        \draw [dashed, -latex, shorten <= 0.2cm] (reftx) -- (\x, -0.4) node (h\i)[pos = 0.8, xshift = \xs, text opacity = 1.0, fill opacity = 0.5, fill = white, align = center] {$h_{\mathrm{ref}, \i}$};
        \node [below = 0.0cm of h\i] {$\tau_{\mathrm{ref}, \i}$};
    }
    \node [anchor = south] at (3.5, 0.5) {AARX antenna elements};
\end{tikzpicture}
    \caption{Illustration of different-length line-of-sight channels between REFTX and AARX array antenna elements for $M = 8$ AARX antennas.}
    \label{fig:reftxchannels}
    \vspace{-0.5cm}
\end{figure}

The channels between REFTX antenna and the $M$ AARX antenna elements may not all be perfectly identical, which is important to consider when performing \ac{STO} and phase correction based on the REFTX broadcast.
In a distributed setup, individual antennas may be separated by distances of several meters or more.
Even under perfect line-of-sight conditions without scatterers, if the REFTX is not located at an equal distance to all AARX antenna elements, the different distances between REFTX antenna and AARX antenna elements will lead to a fixed delay $\tau_{\mathrm{ref}, m}$, phase shift and path loss in the received signal specific to antenna $m$ as illustrated in Fig. \ref{fig:reftxchannels}.
After \ac{STO} correction, which removes the effect of $\tau_{\mathrm{ref}, m}$, the channel experienced by a single subcarrier is approximately frequency-flat.
For subcarrier index $i$, its effect is described only by a channel coefficient $h_{\mathrm{ref}, m, i} \in \mathbb C$.
With transmitted and received symbols $s_{\mathrm{ref}, i}(k) \in \mathbb C$ and $r_{\mathrm{ref}, m, i}(k)  \in \mathbb C$ for subcarrier $i$ at time instance $k$, our channel model is
\begin{equation}
    r_{\mathrm{ref}, m, i}(k) = h_{\mathrm{ref}, m, i} ~ s_{\mathrm{ref}, i}(k).
    \label{eq:refchannelmodel}
\end{equation}
For all antenna elements and subcarriers, the value of $h_{\mathrm{ref}, m, i}$ is unknown but assumed to be constant over time $k$.
In practice, this is ensured by keeping REFTX and AARX at static positions and making sure that the propagation environment does not change.
Due to the small REFTX bandwidth of $2\,\mathrm{MHz}$ and due to the strong propagation path between REFTX and each AARX antenna, the channel can be assumed to be approximately flat over all subcarriers, i.e., we assume
\begin{equation}
    h_{\mathrm{ref}, m} \approx h_{\mathrm{ref}, m, i} ~ \text{for all $i = 0, \ldots, N_\mathrm{sub, ref} - 1$}.
    \label{eq:refchannelflat}
\end{equation}


\section{MOBTX and Channel Coefficient Extraction}
\begin{figure}
    \includegraphics[width=\columnwidth]{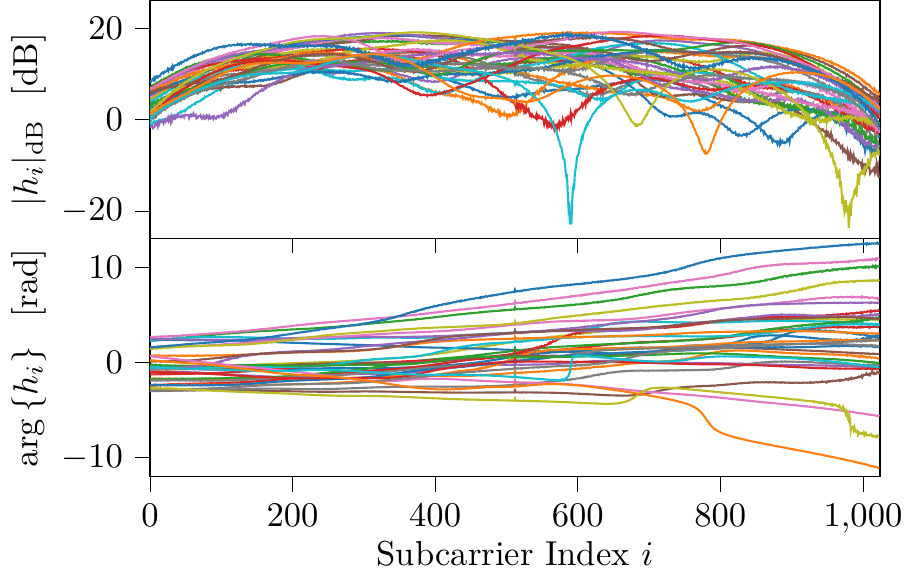}
    \vspace{-0.5cm}
    \caption{Exemplary channel coefficient vectors for an indoor, mostly line of sight-channel with a few clearly visible notches. Each color represents one of the 32 AARX receive antennas. 1024 subcarriers are equally spaced over a bandwidth of $50\,\mathrm{MHz}$.}
    \label{fig:csi}
    \vspace{-0.5cm}
\end{figure}

\label{sec:mobtx}
The channels between MOBTX and each AARX antenna fluctuate as the MOBTX moves in the field.
For the channel between MOBTX and each AARX receive antenna, we use a channel model similar to the one in Eq. \ref{eq:refchannelmodel}, except that the channel coefficients are no longer constant, but dependent on time index $k$:
\begin{equation}
    r_{\mathrm{mob}, m, i}(k) = h_{\mathrm{mob}, m, i}(k) ~ s_{\mathrm{mob}, i}(k)
    \label{eq:mobchannelmodel}
\end{equation}
In Eq. \ref{eq:mobchannelmodel}, $s_{\mathrm{mob}, i}(k)$ denotes the symbol transmitted by the MOBTX at time instance $k$ on subcarrier $i$, $r_{\mathrm{mob}, m, i}(k)$ denotes the corresponding received symbol for the AARX antenna with index $m$ and $h_{\mathrm{mob}, m, i}(k)$ is the time-dependent channel coefficient.
After the sampled received baseband signals for each antenna have been synchronized in time, frequency and phase during postprocessing, subcarrier-specific channel coefficients $h_{\mathrm{mob}, m, i}(k)$ can be extracted in another off-line processing step.
To this end, a perfect reconstruction of the MOBTX transmit signal is produced.
This is achieved by performing a maximal-ratio combining of the MOBTX receive signal of all antennas and then demodulating it.
Through the use of error correction codes and checksums, the received symbols on all \ac{OFDM} subcarriers, even those that are not pilots, can be traced back to their ideal constellation points.
The effect of the channel is then computed based on Eq. \ref{eq:mobchannelmodel} with known $r_{\mathrm{mob}, m, i}(k)$ and $s_{\mathrm{mob}, i}(k)$, for all antennas and all subcarriers.
If the duration of each \ac{OFDM} symbol is much shorter than the channel coherence interval, multiple channel coefficient vectors within the same coherence interval that are captured within the same burst may be averaged to improve the \ac{SNR}.
The extracted \ac{CSI} for a single coherence interval, referred to as a \emph{datapoint}, is tagged with a timestamp and possibly some additional labels such as the position of the MOBTX.
As illustrated in Fig. \ref{fig:csi}, the \ac{CSI} for a particular time instance consists of multiple vectors of channel coefficients, one per AARX receive antenna.
For the $m$\textsuperscript{th} receive antenna, the channel coefficient vector is given by $\mathbf h_{\mathrm{mob}, m} \in \mathbb C^{N_\mathrm{sub, mob}}$, where the $i$\textsuperscript{th} entry, $h_{\mathrm{mob}, m, i}$, contains the estimated channel coefficient for the MOBTX subcarrier with index $i$.

The use of a reference transmitter instead of distributing clock signals has important ramifications on the interpretation of the extracted channel coefficient phases:
All receivers are synchronized in time and phase to their received REFTX broadcast, but, due to the different channels between REFTX and each AARX antenna as expressed in Eq. \ref{eq:refchannelmodel}, they are not necessarily synchronous in absolute terms, i.e., in phase and sampling instances.
The REFTX channel's phase shift leads to a constant, antenna-specific phase rotation.
Moreover, the different propagation path delays between REFTX and AARX antennas lead to a receiver-specific \ac{STO}, which manifests itself as a subcarrier-dependent phase shift that increases or decreases linearly over the subcarrier frequency.

However, by ensuring the propagation environment between REFTX and AARX is static, all phase rotations remain constant and can be accounted for or even be removed through calibration.
In many applications, including precoding for massive \ac{MIMO} and \ac{CSI}-fingerprinting-based positioning, phase differences between antennas that are accurate in absolute terms are not even required, as long as they are correct in relative terms.
That is, phase offsets of individual antennas and subcarriers caused by the REFTX propagation path need to be constant over time and irrespective of the MOBTX position, which is the case.

We denote the REFTX-\emph{adjusted} channel coefficient vector between MOBTX and AARX antenna $m$, i.e., the true channel coefficients that are unaffected by the channel between REFTX and AARX, by $\mathbf a_{\mathrm{mob}, m}$.
Neglecting the propagation delay, the channel between REFTX and AARX antenna $m$ is described by a single channel coefficient $h_{\mathrm{ref}, m}$ (compare Eq. \ref{eq:refchannelflat}), which only leads to a constant phase rotation.
If $h_{\mathrm{ref}, m}$ is known, the adjusted MOBTX channel coefficient vector $\mathbf a_{\mathrm{mob}, m}$ can be computed as
\begin{equation}
    \mathbf a_{\mathrm{mob}, m} = \mathbf h_{\mathrm{mob}, m} \mathrm{e}^{\mathrm{j} \arg\{h_{\mathrm{ref}, m}\}} ~ \text{for $\tau_{\mathrm{ref}, m} \approx 0$}.
    \label{eq:phasecorrection}
\end{equation}
If, as is usually the case, the propagation path delays through the REFTX channel $\tau_{\mathrm{ref}, m}$ are not negligible, the effect of a constant \ac{STO} relative to other AARX receivers also has to be accounted for.
For small \ac{STO} values and narrow subcarrier spacing $\Delta f_\mathrm{sub, mob}$, if $\tau_{\mathrm{ref}, m}$ is known, the effect of \ac{STO} can be accounted for by undoing the phase shift it causes to subcarrier $i$ with $i = 0, \ldots, N_\mathrm{sub, mob} - 1$:
\begin{equation}
    a_{\mathrm{mob}, m, i} = h_{\mathrm{mob}, m, i} ~ \mathrm{e}^{\mathrm{j} (\arg\{h_{\mathrm{ref}, m}\} + 2 \pi i \tau_{\mathrm{ref}, m} \Delta f_\mathrm{sub, mob})}
    \label{eq:stocorrection}
\end{equation}
On the other hand, the amplitudes of the coefficients $\mathbf h_{\mathrm{mob}, m}$ are not influenced by the REFTX propagation path, because measuring them does not require timing synchronization.
In our datasets, all channel coefficient amplitudes are provided relative to an arbitrary, but constant reference amplitude level that is specific to each measurement campaign.

\section{Verification: Phase Stability}
\label{sec:verification}
We carried out a set of experiments to verify that the proposed and implemented system architecture provides accurate and repeatable measurement results.
One particularly insightful experiment demonstrates long-term phase stability and repeatability of \ac{CSI} measurements.
For this experiment, a MOBTX center frequency of $1.27\,\mathrm{GHz}$ with a bandwidth of $8\,\mathrm{MHz}$ and $N_\mathrm{sub, mob} = 1024$ subcarriers was chosen and the REFTX operated at $1.262\,\mathrm{GHz}$.

For a small-scale indoor test setup, the delay spread between multiple propagation paths is very small and the path delay between MOBTX and $m$\textsuperscript{th} AARX antenna element is assumed to only lead to a phase shift.
According to the channel model given in Eq. \ref{eq:mobchannelmodel}, this phase shift is dependent on the subcarrier index $i$.
For illustration purposes, we averaged the phases of all subcarriers, resulting in an average phase shift $\vartheta_m(k)$ perceived by the $m$\textsuperscript{th} antenna at time index k, given by
\begin{equation}
    \vartheta_m(k) = \arg \left\{ \sum_{i = 0}^{N_\mathrm{sub, mob} - 1} h_{\mathrm{mob}, m, i}(k) \right\}.
\end{equation}

\begin{figure}
    \input{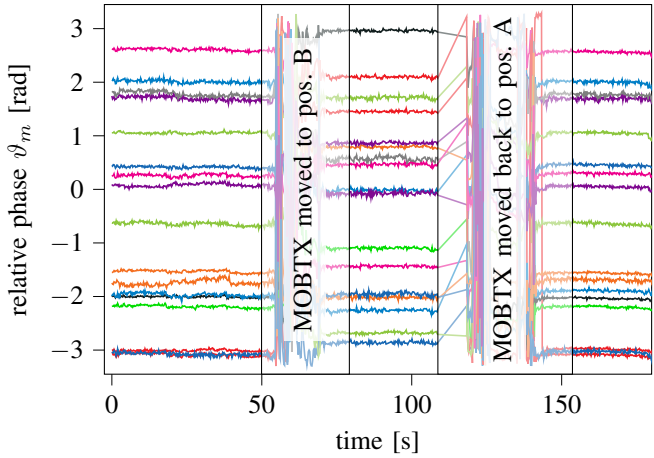}
    \caption{Relative phases $\vartheta_0, \ldots, \vartheta_{16}$ from 17 (out of 32) antennas as measured in the experiment, plotted over time, verifying stability and spatial consistency. The other 15 antennas exhibited similar behavior and are only omitted as to not overload the diagram with too many colors.}
    \label{fig:check}
    \vspace{-0.2cm}
\end{figure}

The measured average phases $\vartheta_m(k)$ of 17 out of 32 antennas, as a function of time elapsed since the start of the experiment, are shown in Fig.~\ref{fig:check}.
In the beginning of the experiment, the MOBTX together with the REFTX and a single, concentrated 32-antenna array as AARX were placed in a static environment (a seminar room).
The initial MOBTX placement is referred to as \emph{Position A}.
Channel measurements are acquired continuously throughout the experiment.
After $50\,\mathrm{s}$ of waiting, the MOBTX is moved to a different position in the same room, referred to as \emph{Position B}.
In Fig.~\ref{fig:check}, the change in received MOBTX phases is clearly visible.
After approximately $30\,\mathrm{s}$ at this new position, the MOBTX is returned to its original location, Position A.
As can be observed from Fig.~\ref{fig:check}, all phases return to their original value, illustrating repeatability and spatial consistency of the set-up.

\section{Dataset Description}
\label{sec:dataset}
At the point of writing, several datasets, measured in both indoor and outdoor environments, have been collected using \ac{DICHASUS}.
As an example that conveys what a typical dataset looks like, some characteristics of one particular indoor dataset will be outlined in the following.
This dataset, called \texttt{dichasus-015x}, was collected using a robot, on top of which the MOBTX was mounted.
The robot was programmed to drive along a meandering path inside certain areas of the room, while \ac{CSI} was acquired by the AARX.
This procedure was repeated multiple times, resulting in \ac{CSI} datasets for multiple \emph{round trips} of the robot, with a duration of 10 to 20 minutes each.
Channel coefficients from $N_\mathrm{sub, mob} = 1024$ subcarriers, spanning a bandwidth of $50\,\mathrm{MHz}$, were acquired.
In total, \ac{CSI} was collected over a duration of $5467\,\mathrm s$ amounting to 77607 datapoints or $20.4\,\mathrm{GB}$ of uncompressed, single precision complex floating point numbers.
On average, more than 14 datapoints were collected per second with each datapoint containing channel coefficients averaged over complete capture bursts containing 90 MOBTX symbols.
In addition to the channel coefficients, each datapoint is tagged with two MOBTX antenna position labels, one derived from a LiDAR that is part of the robot and one generated by a tachymeter tracking the antenna, as well as a timestamp, \ac{SNR} and \ac{CFO} information.

\section{Verification: NN-based Positioning}
\label{sec:nn}

\begin{figure}
    \centering
    \scalebox{0.9}{
    \begin{tikzpicture}
    	\node (l1) [minimum width = 3cm, draw, thick, inner sep = 5pt] at (0, 0) {Flatten};
    	\node (l2) [minimum width = 3cm, draw, thick, inner sep = 5pt, below = 0.6cm of l1] {Dense};
    	\node (l3) [minimum width = 3cm, draw, thick, inner sep = 5pt, below = 0.6cm of l2] {Dense};
    	\node (l4) [minimum width = 6cm, draw, thick, inner sep = 5pt, below = 0.6cm of l3, xshift = 1.5cm] {Dense};
    	\node (l5) [minimum width = 6cm, draw, thick, inner sep = 5pt, below = 0.6cm of l4] {Dense};
    
    	\node [anchor = south] at ($(l1.north) + (0, 0.5)$) {\footnotesize $\left[\,\overline h_{\mathrm{mob}, m, \ell}\,\right]_{m, \ell} \in \mathbb C^{32 \times 64}$};
    	\node [anchor = north] at ($(l5.south) + (0, -0.5)$) {$(\hat x, \hat y)$};
    	\node [anchor = south] at ($(l1.north) + (3.5, 0.5)$) {\footnotesize $\left[\,\overline P_{\mathrm{mob}, \ell}\big|_\mathrm{Np}\,\right]_{\ell}  \in \mathbb R^{64}$};
    
    	\draw [-latex] ($(l1.north) + (0, 0.5)$) -- (l1) node[midway, inner sep = 0] (p0to1) {};
    	\draw [-latex] (l1) -- (l2) node[midway, inner sep = 0] (p1to2) {};
    	\draw [-latex] (l2) -- (l3) node[midway, inner sep = 0] (p2to3) {};
    	\draw [-latex] (l3) -- ($(l4.north) + (-1.5, 0)$) node[midway, inner sep = 0] (p3to4) {};
    	\draw [-latex] (l4) -- (l5) node[midway, inner sep = 0] (p4to5) {};
    	\draw [-latex] (l5) -- ($(l5.south) + (0, -0.5)$) node[midway, inner sep = 0] (p5to6) {};
    	\draw [-latex] ($(l1.north) + (3, 0.5)$) -- ($(l4.north) + (1.5, 0)$) node[midway, inner sep = 0] (p0to4) {};
    
    	\draw [thick] ($(p0to1) + (-0.1, -0.05)$) -- ($(p0to1) + (+0.1, +0.05)$) node[midway, anchor = west, xshift = 0.2cm] {\footnotesize $32 \times 64 \times 2$};	
    	\draw [thick] ($(p1to2) + (-0.1, -0.05)$) -- ($(p1to2) + (+0.1, +0.05)$) node[midway, anchor = west, xshift = 0.2cm] {\footnotesize $4096$};
    	\draw [thick] ($(p2to3) + (-0.1, -0.05)$) -- ($(p2to3) + (+0.1, +0.05)$) node[midway, anchor = west, xshift = 0.2cm] {\footnotesize $512$};
    	\draw [thick] ($(p3to4) + (-0.1, -0.05)$) -- ($(p3to4) + (+0.1, +0.05)$) node[midway, anchor = west, xshift = 0.2cm] {\footnotesize $128$};
    	\draw [thick] ($(p4to5) + (-0.1, -0.05)$) -- ($(p4to5) + (+0.1, +0.05)$) node[midway, anchor = west, xshift = 0.2cm] {\footnotesize $128$};
    	\draw [thick] ($(p5to6) + (-0.1, -0.05)$) -- ($(p5to6) + (+0.1, +0.05)$) node[midway, anchor = west, xshift = 0.2cm] {\footnotesize $2$};
    	\draw [thick] ($(p0to4) + (-0.1, -0.05)$) -- ($(p0to4) + (+0.1, +0.05)$) node[midway, anchor = west, xshift = 0.2cm] {\footnotesize $64$};
    \end{tikzpicture}
}
    \caption{Architecture of the \ac{NN} used for \ac{CSI}-based MOBTX localization.}
    \label{fig:positioning_network}
    \vspace{-0.5cm}
\end{figure}
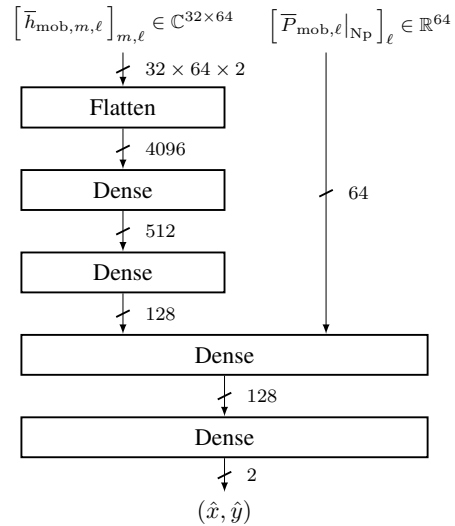

In another experiment, we verified that \ac{CSI} and ground truth labels are suitable for a practical application based on \ac{CSI} data, namely \ac{CSI}-based positioning using \acp{NN}.
We apply multiple robot round trips of aforementioned \texttt{dichasus-015x} dataset to this task.
As a network architecture, the simple \ac{NN} with four dense layers shown in Fig. \ref{fig:positioning_network} is used.
We do not provide the channel coefficients for all 1024 subcarriers directly to the network, but only features derived from them, which are computed in a feature extraction stage.
These features are:
\begin{itemize}
    \item Mean normalized channel coefficients: For each antenna $m = 0, \ldots, M-1$, the channel coefficients of $p$ adjacent subcarriers are summed and then normalized over all antennas $n = 0, \ldots, M - 1$.
    We obtain averaged, normalized channel coefficients with $\ell = 0, \ldots, \sfrac{N_\mathrm{sub, mob}}{p} - 1$:
    \begin{equation}
        \overline h_{\mathrm{mob}, m, \ell} = \frac{\sum_{i = p \ell}^{p\ell + p - 1} h_{\mathrm{mob}, m, i}}{\sqrt{\sum_{n = 0}^{M-1} \left|\sum_{i = p \ell}^{p\ell + p - 1} h_{\mathrm{mob}, n, i}\right|^2}}
        \label{eq:channelcoeff}
    \end{equation}
    The normalization ensures that $\sum_{m = 0}^{M-1} |\overline h_{\mathrm{mob}, m, \ell}|^2 = 1 ~ \forall \ell$.
    As the numeric values of channel coefficients span several orders of magnitude for typical radio environments, this input normalization greatly improves \ac{NN} performance.
    \item Subcarrier powers: The mean power over all antennas of the averaged channel coefficients before the normalization in Eq. \ref{eq:channelcoeff} in logarithmic units:
    \[
        \overline P_{\mathrm{mob}, \ell}\big|_\mathrm{Np} = \ln{\sum_{m = 0}^{M-1} \left|\sum_{i = p\ell}^{p\ell + p - 1} h_{\mathrm{mob}, m, i}\right|^2}
    \]
    The intention of providing these power values is to give the network an indication of the distance between MOBTX and AARX.
\end{itemize}

We choose $p = 16$, yielding $\sfrac{N_\mathrm{sub, mob}}{p} = 64$ averaged normalized channel coefficients for our dataset with 1024 subcarriers.
Since we provide the real and imaginary parts of $\overline h_{\mathrm{mob}, m, \ell} \in \mathbb C$ for all $M = 32$ antennas separately to the \ac{NN}, the averaged normalized channel coefficient input tensor's dimensionality is $M \times \sfrac{N_\mathrm{sub, mob}}{p} \times 2$.

\begin{figure}
    \input{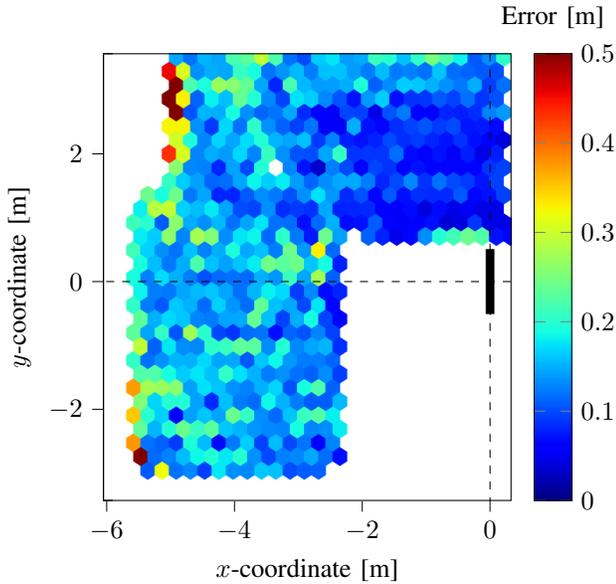}
    \caption{Absolute localization error for an exemplary dataset with concentrated AARX antenna array (black rectangle) at position $(x, y) = (0, 0)$.}
    \label{fig:positioning}
    \vspace{-0.5cm}
\end{figure}


Training is performed on a training set that contains all datapoints from two complete round trips, with batch sizes 32, 64, 256, 1024 and 4096 and 20 epochs per batch size.
The trained network then estimates positions based on \ac{CSI} from a third round trip, our test set.
While the robot always stays inside the same part of the room during every round trip, it may drive along a different path, hence the visited locations from one round trip do not precisely match those from previous trips.
By training and evaluating on different robot round trips, the network is presented inputs from positions in the room that it has never seen before during training.
Fig. \ref{fig:positioning} illustrates the quality of the position estimates $(\hat x, \hat y)$ generated by the \ac{NN} for the outlined scenario.
Over the complete test set, we achieve a mean absolute error of less than $0.2\,\mathrm{m}$.

\section{Summary and Outlook}
The system architecture of a distributed massive \ac{MIMO} channel sounder with over-the-air synchronization, suitable for researching cell-less massive \ac{MIMO} systems and various antenna deployments, has been presented.
To verify the suitability of our approach, first captured datasets have been generated and analyzed.
The captured datasets can be used to tackle open research questions in a variety of domains and to empirically verify theoretical results.


Two particularly interesting applications are channel charting \cite{studer} and enabling massive MIMO in \ac{FDD} operation \cite{arnold2019enabling}.
\ac{CSI} captured by \ac{DICHASUS} may be used to verify results previously only obtained through simulations with real-world measurement data.

\bibliographystyle{IEEEtran}
\bibliography{IEEEabrv,references}

\end{document}